\documentclass[aps,prd,superscriptaddress,nofootinbib,twocolumn]{revtex4}

\pdfoutput=1

\usepackage{amsfonts,amsmath,amsthm,amssymb}
\usepackage{mathrsfs}
\usepackage{bm}
\usepackage{color}
\usepackage{epsfig}

\usepackage[usenames,dvipsnames]{xcolor}
\definecolor{orange}{cmyk}{0,0.5,1,0}
\definecolor{rossoCP3}{cmyk}{0,.88,.77,.40}
\definecolor{graa}{rgb}{0.8,0.8,0.8}
\definecolor{blaa}{rgb}{0.2,0.2,0.6}

\newcommand{\PRE}[1]{{#1}}   
\newcommand{\met} {\not\!\! E_T}

\newcommand{\comment}[1]{}

\newcommand{\beq}[1]{\begin{equation}\label{#1}}
\newcommand{\eeq}{\end{equation}}
\newcommand{\be}{\begin{equation}}
\newcommand{\ee}{\end{equation}}
\newcommand{\beqa}[1]{\begin{eqnarray} \label{#1}}
\newcommand{\eeqa}{\end{eqnarray}}
\newcommand{\bea}{\begin{eqnarray}}
\newcommand{\eea}{\end{eqnarray}}
\newcommand{\bay}[1]{\left(\begin{array}{#1}}
\newcommand{\eay}{\end{array}\right)}
\newcommand{\ba}{\begin{array}}
\newcommand{\ea}{\end{array}}
\newcommand{\rf}[1]{(\ref{#1})}

\newcommand{\gs}{\mathrel{
   \rlap{\raise 0.511ex \hbox{$>$}}{\lower 0.511ex \hbox{$\sim$}}}}
\newcommand{\ls}{\mathrel{
   \rlap{\raise 0.511ex \hbox{$<$}}{\lower 0.511ex \hbox{$\sim$}}}}
\newcommand{\Ecut}{E^\nu_{\max}}

\newcommand{\nue}{{\nu_{e}}}
\newcommand{\numu}{{\nu_{\mu}}}
\newcommand{\nutau}{{\nu_{\tau}}}
\newcommand{\nuebar}{{\bar \nu}_e}

\def\met{\mbox{${\hbox{$E$\kern-0.6em\lower-.1ex\hbox{/}}}_T$}}   

\begin{document}

\title{\PRE{\vspace*{0.9in}} \color{rossoCP3}
{
A Relational Argument for a \boldmath{$\sim$}PeV Neutrino Energy Cutoff
}
	
\PRE{\vspace*{0.1in}} }





\author{J.G.~Learned}
\affiliation{Department of Physics \& Astronomy,\\
University of Hawaii at Manoa, Honolulu, HI 96822, USA
\PRE{\vspace*{.1in}}
}




\author{T.J.~Weiler}
\affiliation{Department of Physics \& Astronomy,\\
Vanderbilt University, Nashville TN 37235, USA
\PRE{\vspace*{.1in}}
}


\begin{abstract}
 \PRE{\vspace*{.1in}} 
\noindent 
We present a relationship among the highest observed neutrino energy ($\sim$PeV)
and the neutrino mass, the weak scale, and the Planck energy:
$\Ecut = \frac{m_\nu\,M_{\rm Planck}}{M_{\rm weak}}$.
We then discuss some tests of this relationship, 
and present some theoretical constructs which motivate the relationship.
It is possible that all massive particles are subject to maximum energies given by 
similar relationships, although only the neutrino seems able to offer 
interesting phenomenology. We discuss implications which include no neutrino detections at energies greater than PeV, 
and changes in expectations for the highest energy cosmic rays.  A virtue of this hypothesis is that it is easily invalidated 
should neutrinos be observed with energies much great than the PeV scale. An almost inescapable implication is that 
Lorentz Invariance is a low energy principle, yet it appears that violation may be only observable in high-energy astrophysical neutrinos.

%
\end{abstract}
\pacs{xxxx}

\maketitle

\section{Introduction}
\label{sec:intro}
In 2012, IceCube released a two-year equivalent data set,
observing for the first time high-energy non-atmospheric
neutrino events~\cite{Aartsen:2013bka,Aartsen:2013jdh}.
The maximum neutrino energy was inferred to be $\sim$PeV\@.
In 2014, IceCube reported its 3yr data set~\cite{IceCube3yr}.
In the 3yr sample, the maximum neutrino energy was inferred to be $\sim 2$~PeV.
The energy resolution on the observed events is $\sim 25$\%.
The IceCube experiment itself has commented on the apparent cutoff on the neutrino energy.
Further credence for a neutrino cutoff energy arises from the fact that ``Glashow resonance"~\cite{Glashow:1960zz} 
events (${\bar \nu_e} +e^- \rightarrow W^- \rightarrow {\rm shower}$ at $E_{\bar \nu_e}=6.3$~PeV) are not yet observed, 
even though the effective area for such events is relatively large (and of course, dependent on the $\nuebar$ flux at Earth)~\cite{GRes}.
And even further credence for a cutoff in neutrino energy comes form the non-observation of any higher energy neutrino events in 
the ANITA~\cite{ANITA}, RICE~\cite{RICE}, Auger~\cite{Auger}, HiRes~\cite{HiRes}, and Telescope Array~\cite{TA} experiments, plus the reported non-observation of higher energy traversing muons in IceCube. 

\section{The Relationship}
\label{sec:relationship}
In this Letter we take the evidence for a cutoff in neutrino energy seriously 
(as was done in a strictly phenomenological way in Ref.~\cite{EnuCutoff}).
We note that a simple relationship among known energy and mass scales leads to 
just the maximum neutrino energy observed at IceCube.
The relationship is 
\bea
\label{relation}
&& \Ecut = \frac{m_\nu\,M_{\rm Planck}}{M_{\rm weak}}\, \\
&& = 1.2 \left( \frac{m_\nu}{0.1\,{\rm eV}} \right)
\left( \frac{M_{\rm Planck}}{1.2\times 10^{28}\,{\rm eV}} \right)
\left( \frac{100\,{\rm GeV}}{M_{\rm weak}} \right) {\rm PeV}\,. \nonumber
\eea
In the above relationship, the neutrino mass, Planck mass, and weak scale are each 
scaled by expected values.\footnote
{The factor $1.2\left( \frac{100\,{\rm GeV}}{M_{\rm weak}} \right)$ may be replaced by 
$1.0\left( \frac{126\,{\rm GeV}}{m_{\rm Higgs}} \right)$ or by $0.5\left( \frac{247\,{\rm GeV}}{v_{\rm weak}} \right)$.
The point is that the weak scale enters here.
}

The variables on the right-hand side of this equation are numbers, Lorentz scalars;
the variable in the left-hand side is the zeroth component of a four-vector, not  a Lorentz invariant.
Thus one must address in what frame Eq.~\rf{relation} is expected to hold.
Nature has supplied a preferred frame, the cosmic rest frame (CRF).  
In the CRF, the CMB temperature is homogeneous and well-defined, at 2.73K.  
The four-velocity of the CRF is $u^\alpha_{\rm CRF}\equiv \left( \frac{dx}{d\tau}\right)_{\rm CRF}=(1,\vec{0})$.
Our Earthly rest frame is removed from this frame by a non-relativistic boost of 370~km/s,
as evidenced in an earthly observation of a dipole component of the CMB.
Since this boost velocity to the CRF is just 0.12\% times the speed of light,
we may neglect this boost in what follows, and take our Earth-frame and CRF to be one and the same.
Consequently, the maximum neutrino energy in the Earth frame is also the maximum energy in the CRF.
We may write this maximum energy as a Lorentz invariant,
$\Ecut=P^\nu_\alpha u^\alpha_{\rm CRF}$, so that all terms in Eq.~\rf{relation} are now Lorentz scalars.

Eq.~\rf{relation} may be construed as the centerpiece of this paper.

\subsection{Tests and Signatures}
\label{subsec:TandS}
This relationship~\rf{relation} can be easily disproven experimentally. 
If neutrino events at energies a factor of a few above the present maximum are observed, 
then the relationship is dead.
The relationship will also be disproven if cosmogenic neutrinos~\cite{BZ}, expected in the SM 
with energies peaking at $\sim 10^{18}$~eV, are observed.

The linear dependence of $\Ecut$ on neutrino mass has implications for the 
track to shower ratio of observed neutrino events at their highest observable energies.
In the normal hierarchy, $\nu_3$ is the heaviest neutrino, with flavor content known to be 
roughly $(\nue$:$\numu$:$\nutau)=$(0:1:1);
in the inverted hierarchy, $\nu_2$ and $\nu_1$ are the heavier neutrinos, with collective flavor content 
known to be (2:1:1).  Thus, a prediction inherent in the relation presented in Eq.~\rf{relation} is that 
if the neutrino masses are hierarchical (meaning, having distinctly different masses), 
then the track-to-shower ratio at Earth will approach $1:1$ for the normal hierarchy and $1:3$ for the inverted hierarchy
at the highest neutrino energies.\footnote
{We note the contrast with neutrino decay models~\cite{NuDK}, for which it is the flavor content of the lightest neutrinos 
that determine flavor ratios (e.g. the track-to-shower ratio) at Earth.  Here it is the heavier neutrinos which 
determine the high-energy flavor content.
} 
(We have ignored the lower-energy neutral current events, 
which are insignificant within a spectrum falling as a power law.)

Next we present a physical argument that is motivated by the relationship in Eq.~\rf{relation}.\footnote
{The fact is that we constructed the physical argument for the neutrino cutoff energy {\bf before} 
we found the simple relationship in Eq.~\rf{relation}.  
}

\subsection{Theoretical Motivation}
\label{subsec:motivation}
Eq.~\rf{relation} can be re-written as 
\beq{relation2}
\Gamma_\nu^{\max}\,M_{\rm weak} = M_{\rm Planck}\,,
\eeq
where $\Gamma_\nu \equiv E_\nu/m_\nu$ is the usual Lorentz factor.

We note that the same relational equation may apply for all particles with mass.
The next-lightest known particle to the neutrino is the electron, 
for which the cutoff energy according to Eq.~\rf{relation} would be $\sim 6\times 10^{21}$~eV 
well above what has ever been observed for an electron, 
and far above what can be measured since the light, charged electron at such an initial energy 
would suffer severe and immediate synchrotron losses.  
For the muon, pion, and nucleon, we get respective maximum energies from Eq.~\rf{relation} in the 
$10^{24}$ to $10^{25}$~eV range, far beyond what has been measured in cosmic ray physics.
Of course, all fermions beyond the neutrino have interactions stronger than weak.
Accordingly, it may be that the Eq.~\rf{relation2}, with the weak scale explicitly exhibited on the left-hand side,
only pertains to the neutrino.

We show in Fig.~\rf{fig:nuend} the relationships among $\Gamma^{\max}$ and maximum energies for some light particles
and the Higgs boson.
Also shown, for scale, are the Glashow resonance energy, the apparent end of the cosmic-ray spectrum,
the GUT energy, and the Planck energy.
Notice that $\Gamma_\nu$ crosses $M_{\rm Planck}/M_{\rm weak}$ at $E_\nu\sim$PeV,
the core observation of this Letter.
\begin{figure}
\centering
\includegraphics[width=0.55\textwidth]{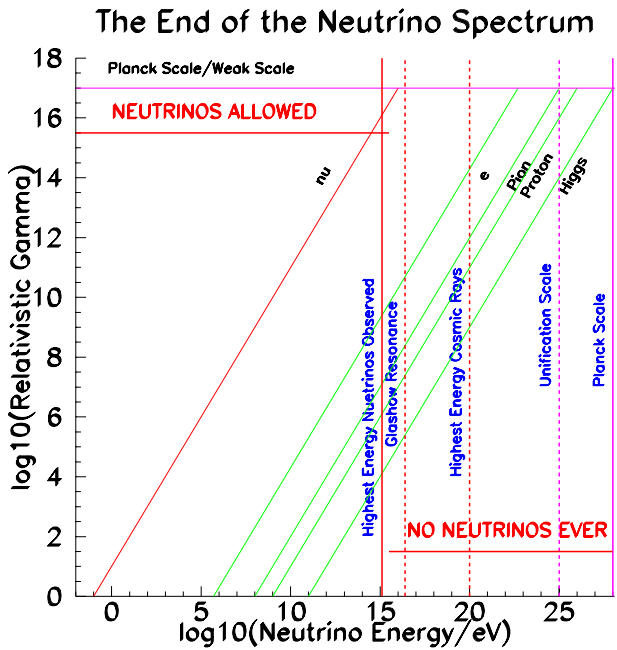}
\caption{Shown are the relationships among $\Gamma^{\max}$ and maximum energies for the neutrino ($0.1$~eV mass assumed), 
electron, pion, proton, and Higgs boson.
Also shown, with vertical lines, are other significant high-energy scales:
the Glashow resonance energy, the apparent end of the cosmic-ray spectrum,
the GUT energy, and the Planck energy.
Notice that $\Gamma_\nu$ crosses $M_{\rm Planck}/M_{\rm weak}$ at $E_\nu\sim$PeV,
thereby establishing $\Ecut\sim m_\nu \frac{M_{\rm Planck}}{M_{\rm weak}}\sim$PeV as the 
maximum neutrino energy.}
\label{fig:nuend}
\end{figure}

One approach to interpreting Eq.~\rf{relation2} is simply to say that the Lorentz factor 
cuts off at a scale given by the ratio of the two important high-energy mass scales, the weak and the Planck:
$\Gamma=\frac{M_{\rm {Planck}}}{M_{\rm weak}} \sim 10^{17}$.  
There is no reason known to us why this cutoff must occur in present standard theory.
On the other hand, before the recent IceCube observation of extraterrestrial neutrinos there has never been an observable particle whose $\Gamma$~factor could even approach this enormous but finite value. In this picture, it seems inescapable that Lorentz Invariance is an emergent symmetry, valid for ``low''~$\Gamma$.~\cite{emergentLI}

A more intriguing interpretation of Eq.~\rf{relation2} contains dynamics.
If one postulates that the characteristic ``size'' of the neutrino in the rest frame 
of the decaying parent particle (charged pion, muon, charmed particle, etc.)
is of order of the inverse weak scale $1/M_{\rm weak}$.\footnote
{
The inverse weak scale $(100\,{\rm GeV})^{-1}$ corresponds to $2\times 10^{-3}\,{\rm fm}$.
}
Then, according to Special Relativity, in the lab frame the neutrino size is spatially contracted to $1/(\Gamma_\nu\,M_{\rm weak})$.
When this contracted size is of order of a Planck length $1/M_{\rm Planck}$,
a new phenomenon takes place which effectively removes the neutrino from observation.
Here it is the Planck scale that initiates the new physics, not an unusual expectation, though one that has not been considered for neutrinos heretofore.

Alternatively, one may postulate that the weak interaction is the characteristic size of space-time foam
(perhaps due to fluctuations in the Higgs vacuum).  
The characteristic size of space-time as seen by a relativistic neutrino is then 
$1/(\Gamma_\nu\,M_{\rm weak})\sim 1/M_{\rm Planck}$.
Again, it is the Planck scale that initiates the new physics.

This new phenomenon at the Planck scale may take several forms.
Some possibilities are \\
(i) Gravity becomes strong for the neutrinos at the Planck scale, either preventing the formation of the neutrino wave packet or
 presenting a strong cross section for neutrino scattering off gravity/geometry, with significant loss of neutrino energy. \\
(ii) Space-time manifests itself as foam at the Planck scale, either preventing the formation of the neutrino wave packet or
 presenting a strong neutrino-foam scattering cross-section with significant loss of neutrino energy. 
 These continued foam interactions are reminiscent of the quantum Zeno effect.\\
(iii) Lorentz Invariance is violated (LIV) at the Planck scale.
A simple manifestation of LIV, broken rotational symmetry,  results if space dimensions are latticized at the Planck scale, 
as often discussed over the past decades; 
in the present context, the manifestation of LIV is the apparent maximum neutrino energy.\\
(iv) The neutrino may even transit from our brane into extra space dimension(s) having the scale size 
natural to gravity, the Planck length.

More illustrations can be constructed.

What these four illustrative examples have in common is that when the neutrino size becomes of order of the Planck length,
some new phenomenon occurs that prevents decay into a neutrino, or removes the neutrino itself from the observable realm.
This phenomenon is summarized mathematically in our Eq.~\rf{relation}.

It is desirable to propose tests that discriminate among various models of the new phenomenon,
though it seems premature to do so in any detail.
The whole class of such models first needs verification of a continued cutoff at $\sim$PeV
in the face of improved statistics.
Still, there arguments presented above for a maximum neutrino energy fall into one of two classes:
either the neutrino can be produced with $E_\nu > \Ecut$ and subsequently lose energy,
or the neutrino can only be produced with $E_\nu\le \Ecut$.
There is a marked difference between the two classes, leading to possible discrimination.
In the former case, there will be a mild pileup of neutrino events in the bin of size $\Delta E$ at $\Ecut$, 
given by the integral contribution above $\Ecut$, i.e., $\frac{\Ecut}{\alpha-1}$, where $\alpha$,
assuming a power law, is the spectral index continued above $\Ecut$.
Also, in the latter case, but not in the former case, high-energy decays of parent particles to final states that 
otherwise would contain neutrinos are suppressed; 
this increases the parent's lifetime, or even stabilizes the parent 
(as has been noted recently for the pion parent~\cite{EnuCutoff}).

Perhaps the weakest link in our model construction of an attainable Planck length
when a neutrino is sufficiently Lorentz-boosted, 
is the association of the unboosted ``size'' with the inverse weak scale.
Still, it is not unreasonable that the neutrino at production should somehow be intimately 
aware of the weak-interaction scale.  
Alternatively, it does not seem unreasonable that fluctuations in spacetime
may have a characteristic scale given by the inverse of the Higgs vev (247~GeV)
or Higgs mass (126~GeV).

\section{Discussion and Summary}
\label{sec:DandS}
We have presented in Eq.~\rf{relation} a relationship among known quantities that results in a new neutrino scale of a~PeV\@.
Faced with the recent IceCube observations of astrophysical neutrinos with energies up to $\sim2$~PeV but none beyond, 
nor any other neutrino observations from other experiments probing higher energies,
it is tempting to associate this new scale with a maximum observable energy for neutrinos in our 4D Universe, $\Ecut$.

We have offered sketches for models underpinning this $\Ecut$-relationship.
One construction is to simply admit that a maximum $\Gamma_\nu$ has been discovered,
equal to the ratio $\frac{M_{\rm Planck}}{M_{\rm weak}}\sim 10^{17}$.
Another construction postulates a fundamental ``size'' given by the Lorentz-contracted inverse weak-scale.
We offered two possible assignments for this size, one associated with the highly-relativistic, weakly-interacting neutrino,
the other with fluctuations in space-time of the Higgs vev as seen by the neutrino.
We noted that at a $\Gamma$ of $\frac{M_{\rm Planck}}{M_{\rm weak}}\sim 10^{17}$, 
a number natural for neutrinos at $~$PeV,
the Lorentz-contracted size is equal to the Planck length.
Planck-scale physics then offers a multiplicity of possibilities for cutting off the neutrino energy. 
As illustration, we listed four possibilities, each of which would motivate Eq.~\rf{relation}. 
And there may be other constructions beyond the ones we listed that are even ``better'', 
in the sense of being closer to Nature's reality.

We have outlined a few observable consequences of our linear relation between maximum observable energy and particle mass.
Should this relationship hold, it may herald an opening to exploration of physics at the Planck scale.
For all massive particles but the neutrino, the relation appears to imply no new observable phenomenon.
However, for the neutrino the relation implies, in addition to a $\sim$PeV cutoff energy, 
some possible new phenomenology.
If the neutrino masses are hierarchical so that $\Ecut$ depends significantly on flavor,
then Interesting predictions for the track-to-shower event ratio at the highest neutrino energies emerge.
Also, if neutrinos cannot be produced above $\Ecut$,
then leptonic and semi-leptonic decays of mesons are suppressed above some energy;
the charged pion may even become stable~\cite{EnuCutoff}.

Of course, it must be emphasized that the hypothesis presented herein 
will be invalidated if neutrinos are observed with energies exceeding a few~PeV.

%
\section*{Acknowledgments}
%
JGL 
is  supported by DoE Grant No.\ DE-FG02-04ER41291, and 
TJW is supported by DoE Grant No.\ DE-FG05-85ER40226 and the Simons Foundation Grant No.\ 306329.



\begin{thebibliography}{99}

\bibitem{Aartsen:2013bka} 
  M.~G.~Aartsen {\it et al.}  [IceCube Collaboration],
  Phys.\ Rev.\ Lett.\  {\bf 111}, 021103 (2013)
  [arXiv:1304.5356 [astro-ph.HE]];

\bibitem{Aartsen:2013jdh} 
 M.~G.~Aartsen {\it et al.}  [IceCube Collaboration],
  Science {\bf 342}, no. 6161, 1242856 (2013)
  [arXiv:1311.5238 [astro-ph.HE]].

\bibitem{IceCube3yr}
  M.~G.~Aartsen {\it et al.}  [IceCube Collaboration],
  arXiv:1405.5303 [astro-ph.HE].
  
\bibitem{Glashow:1960zz} 
  S.~L.~Glashow,
  Phys.\ Rev.\  {\bf 118}, 316 (1960).
  
\bibitem{GRes}
  A.~Bhattacharya, R.~Gandhi, W.~Rodejohann and A.~Watanabe,
  arXiv:1209.2422 [hep-ph];
  
  A.~Bhattacharya, R.~Gandhi, W.~Rodejohann and A.~Watanabe,
  JCAP {\bf 1110}, 017 (2011)
  [arXiv:1108.3163 [astro-ph.HE]];
  ibid.,  
  arXiv:1209.2422 [hep-ph].

  M.~D.~Kistler, T.~Stanev and H.~Yuksel,
  arXiv:1301.1703 [astro-ph.HE];

V. Barger, L. Fu, J.G. Learned, D. Marfatia, S. Pakvasa, and T.J.Weiler, 
in progress.


\bibitem{ANITA}
 P.~W.~Gorham {\it et al.}  [ANITA Collaboration],
  Phys. Rev. D {\bf 82} 022004 (2010)
 [arXiv:1003.2961 [astro-ph.HE]];
  
ibid., Erratum: 
  Phys.\ Rev.\ D {\bf 85}, 049901 (2012)
  [arXiv:1011.5004 [astro-ph.HE]].

\bibitem{RICE}
Ilya Kravchenko, Dave Seckel, Dave Besson, John Ralston, John Taylor, Ken Ratzlaff, Rob Young,
 Phys. Rev. D. {\bf 85} 062004 (2011.

\bibitem{Auger}
  P.~Abreu {\it et al.} [Pierre Auger Collaboration],
  Phys.\ Rev.\ D {\bf 84}, 122005 (2011)
  [arXiv:1202.1493 [astro-ph.HE]];
  
 P.~Abreu {\it et al.}  [Pierre Auger Collaboration],
  Astrophys.\ J.\  {\bf 755}, L4 (2012)
  [arXiv:1210.3143 [astro-ph.HE]].

\bibitem{HiRes}
P. Sokolsky; for the HiRes Collaboration (2010). 
arXiv:1010.2690v1 [astro-ph.HE

\bibitem{TA}
H. Tokuno, {\it et al},
Nuclear Instruments and Methods in Physics Research A 676: 54–65 (2012). arXiv:1201.0002

\bibitem{EnuCutoff}
  L.~A.~Anchordoqui, V.~Barger, H.~Goldberg, J.~G.~Learned, D.~Marfatia, S.~Pakvasa, T.~C.~Paul and T.~J.~Weiler,
  arXiv:1404.0622 [hep-ph].

\bibitem{BZ}
  V.~S.~Berezinsky and G.~T.~Zatsepin,
  Phys.\ Lett.\ B {\bf 28}, 423 (1969);

  R.~Engel, D.~Seckel and T.~Stanev,
  Phys.\ Rev.\ D {\bf 64}, 093010 (2001)
  [astro-ph/0101216].
  

\bibitem{NuDK}
  S.~Pakvasa,
  Lett.\ Nuovo Cim.\  {\bf 31}, 497 (1981);
  
 J.~F.~Beacom, N.~F.~Bell, D.~Hooper, S.~Pakvasa and T.~J.~Weiler,
  Phys.\ Rev.\ Lett.\  {\bf 90}, 181301 (2003)
  [hep-ph/0211305];
ibid., 
  Phys.\ Rev.\ D {\bf 69}, 017303 (2004)
  [hep-ph/0309267];
  
  P.~Baerwald, M.~Bustamante and W.~Winter,
  JCAP {\bf 1210}, 020 (2012)
  [arXiv:1208.4600 [astro-ph.CO]].
  
  \bibitem{emergentLI}
  A recent example of Lorentz Invariance emerging at low energies (relative to the Planck mass)
  is given in 
  J.~Khoury, G.~E.~J.~Miller and A.~J.~Tolley,
  arXiv:1405.5219 [gr-qc],
written for the Gravity Research Foundation (Honorable Mention 2014).

\end{thebibliography}
\end{document}